\documentclass[floatfix,showpacs,showkeys,prl,twocolumn]{revtex4}
\usepackage{epsfig}
\usepackage{graphicx}
\usepackage{amssymb}
\usepackage{amsmath}
\usepackage{hyperref}
\usepackage{color}

\begin{document}


\title{Understanding Ultracold Neutron Production in Oxygen Solids from Volume and Temperature Dependent Yields}
\date{\today}

\author{C.-Y. Liu}
\email{CL21@indiana.edu}
\author{C.M. Lavelle}
\author{D.J. Salvat}
\author{W. Fox}
\author{G. Manus}
\author{P.M. McChesney}
\author{Y. Shin}
\affiliation{Physics Department, Indiana University, Bloomington, IN 47408}

\author{A. Couture}
\affiliation{LANSCE Division, Nuclear Science, Los Alamos National Laboratory, Los Alamos, New Mexico 87544, USA}

\begin{abstract}
We present experimental data on ultracold neutron (UCN) production in solid oxygen from 5~K to 50~K. The experiment used 3 cells of different dimensions in order to study the excess UCN elastic scattering in s-O$_2$ due to source density non-uniformities. We present a combined $\chi^2$ analysis with the aid of Monte-Carlo simulations which contain the detailed physics of UCN transport as well as cold neutron attenuation. 
The results validate the UCN production and upscattering cross-sections extracted from the dynamic structure function derived from inelastic neutron scattering, and illustrate the role of magnetic scattering in enhancing UCN production in s-O$_2$.
\end{abstract}



\maketitle


As superthermal ultracold neutron (UCN) sources are currently under construction around the world, using either deuterium or superfluid helium as converter materials, the investigation of other materials is a potentially interesting topic to a broad scientific community. In particular, oxygen, from a theoretical point of view, seems to be a promising alternative, as it has a lower nuclear absorption (compared to deuterium), is commercially available with high purities, and has sufficiently low Fermi-potential, although its UCN production cross section is slightly smaller than that of deuterium. 
Detailed study of this new material could yield new scientific insight.
Unfortunately, experimental data presented in the past few years~\cite{Kasprzak08, Atchison2009, Frei2010} are inconclusive and even contradictory. Despite the fact that these experiments were carefully designed to provide direct comparison of UCN production from the materials under study, interpretation of the results was difficult, mainly because of the intrinsically different thermal properties of these cryogenic solids that limit the knowledge of UCN extraction.    

In a typical UCN production experiment, the observed UCN count rates are a convolution of (a) UCN production, (b) UCN loss during extraction from the source, (c) UCN transport from the source to the detector, and (d) the variation of the detector efficiency due to the different UCN spectrum as a result of the velocity boost that all UCN acquire upon exiting the source material into a vacuum guide tube. The UCN transport can be simulated using detailed Monte-Carlo tools, given that the physics of UCN transport in a vacuum guide is known relatively well. The loss of UCN in the process of extraction from the production source, on the other hand, is highly dependent on the physical condition of the source. Any variations in material density due to cracks or voids can significantly change the UCN mean free path (mfp). Even though visual inspections were frequently made during experiments, most assessments of the solids grown {\it in-situ} are qualitative at best and might not be directly applicable to UCN, due to the fact that the wavelength of UCN is an order of magnitude shorter than that of optical photons. 

Furthermore, previous analyses often lack sufficient treatment of the evolution of the incident cold neutron (CN) flux to reliably extract the physics of UCN production in the source material under study. To better understand the UCN production mechanism, we have decoupled the CN production from the UCN production by placing the experiment on a CN beam line, even though the experiment takes a significant cut in the available CN flux. 
The UCN production source is often made thick in order to increase the probability of downscattering a CN into a UCN. 
In such a source, we expect that CN experience strong elastic Bragg scatterings that are highly energy dependent. Unfortunately, this piece of physics is often neglected, most likely because the standard neutron transport codes (e.g., MCNP~\cite{mcnp5}) do not contain readily available data files that include coherent scattering~\cite{MacFarlane94}. 
There remain many difficulties in ascertaining coherent understanding among many UCN production experiments. 
In this letter, we present an analysis of UCN production data in solid oxygen (s-O$_2$), that attempts to take into account of all the effects mentioned above. Experimental data were taken using three cells of different dimensions to give us a handle on the mfp of UCN. Combined with additional data on inelastic neutron scattering, we demonstrate how the analysis sheds light on the mechanism of UCN production in s-O$_2$ in low temperature phases at the saturated vapor pressure. 


The experimental data presented here were taken at Lujan Center neutron flight path 12 at LANSCE during the summer of 2008. Details of the experimental setup was reported in ~\cite{Lavelle2010}. 
The performance of the apparatus was bench-marked using solid ortho-deuterium 
in the target cell maintained between 5~K and 22~K. 
Because of the low saturated vapor pressure of s-O$_2$, vapor deposition techniques used to attain optically transparent solid deuterium~\cite{morris2002} do not work for oxygen. Instead, gaseous oxygen (at 99.999\% purity) was first condensed into liquid around 60~K in the target cell, and then slowly cooled into solid phases. 
Unlike deuterium, oxygen has three distinct solid phases at saturated vapor pressure: $\gamma$ (44 $\sim$ 55~K, cubic lattice, paramagnetic), $\beta$ (24 $\sim$ 44~K, rhombohedral lattice, short range anti-ferromagnetic(AF)), and $\alpha$ phase ( $<$ 24~K, monoclinic lattice, 2-D AF). 
The strong magnetic interaction between O$_2$ molecules (each with a non-zero electronic spin $S=1$) plays a significant role in the abrupt changes of lattice structures among these low temperature phases. 
We use Bragg edges in the total cross section (obtained from CN transmission data) to verify these lattice structures.

\begin{figure}[t]
\centering
\includegraphics[width=3.3 in]{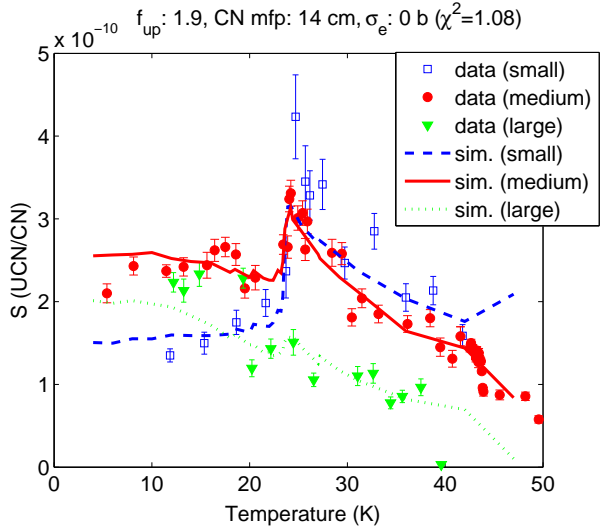} 
\caption{\label{fig:BestFit} Best combined fit of the UCN yields from solid O$_2$ in cells of three different lengths.}
\end{figure}

UCN production was studied using three target cells of different length (small: 1.14 cm, medium: 3.56 cm, large: 8.61 cm), all with the same inner diameter of 6.40 cm. The background subtracted UCN production signals, S, (the UCN count rate normalized to the incident CN current as defined in~\cite{Lavelle2010}) from the three cells are plotted in Fig.~\ref{fig:BestFit}. 
The target cell was cooled slowly from liquid at 60~K, through $\gamma$, $\beta$, and into $\alpha$-O$_2$ at 5~K over the duration of 5 days (following the recipe in~\cite{jez1993}). The major challenge arises from the reduction of the molar volume of s-O$_2$ by 12.5\% upon cooling through the three solid phases at saturated vapor presure. A discrete 5\% change at the $\gamma-\beta$ transition at 44~K is the bottleneck to attaining large-sized s-O$_2$ cryocrystals in the $\beta$ and $\alpha$ phases. 
Even with our best efforts in temperature control, subsequent faster warm-up and cool-down cycles (over 10 hours) did not show any noticeable changes in the UCN production.  
At the first glance, the data presented in Fig.~\ref{fig:BestFit} beg for explanations for (a) their non-superthermal behavior, (b) their non-linear volume dependence, and (c) the peculiar production peak at the $\beta-\alpha$ transition. 
The remainder of the paper is directed to explaining these atypical behaviors.  

\begin{figure}[b]
\centering
\includegraphics[width=3.3 in]{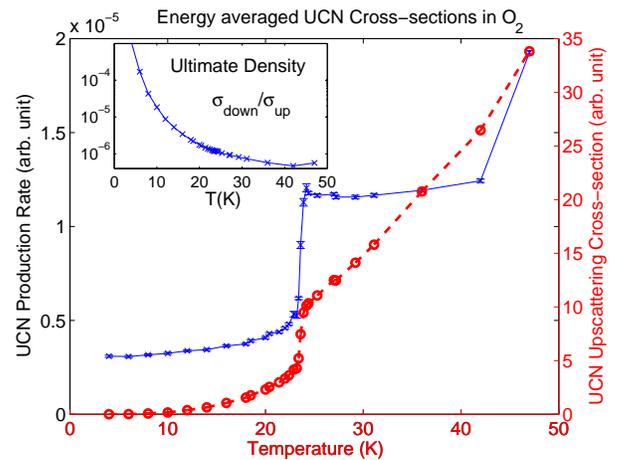}
\caption{\label{fig:UCNXection} Temperature dependent UCN production rate and UCN upscattering cross section in s-O$_2$. Inset: The ultimate achievable UCN density by taking $\sigma_{down}/\sigma_{up}$.}
\end{figure}

Due to the distinct magnetic orderings, we expect the magnon excitations to vary among the three solid phases, even without detailed {\it a priori} understanding of the underlying physics. A follow-up measurement~\cite{Liu2011} of the dynamic structure function, $S(|Q|,\omega)$, of s-O$_2$ was carried out using the Disk Chopper Spectrometer at the NIST neutron scattering center in 2009. 
As described in ~\cite{Liu2011}, we can extract the differential cross section for UCN production and the upscattering using $S(Q,\omega)$, regardless of the properties of the modes of excitation, i.e., 
\begin{eqnarray}
\frac{d\sigma^{down}}{dE} &=& 4\pi \frac{k_{ucn}}{Q} S\left(Q,\omega=\frac{\hbar^2Q^2}{2m_n}\right)_{\omega>0}, \label{Eq:UCNProduction} \\
\frac{d\sigma^{up}}{dE} &=& 4\pi \frac{Q}{k_{ucn}} S\left(Q,\omega=-\frac{\hbar^2Q^2}{2m_n}\right)_{\omega<0},
\label{Eq:UCNup}
\end{eqnarray}
where $\omega=E-E_{ucn}$ is the energy transfer, and $m_n$ is the mass of a free neutron. Integrating Eq.~\ref{Eq:UCNProduction} over the incident CN flux, we then get the spectrum-averaged UCN production rate; integrating Eq.~\ref{Eq:UCNup} over all final phase space, we get the UCN upscattering cross section. 
Tables of $S(Q,\omega)$ were measured at 28 temperature points between 5~K and 47~K. The extracted UCN production rate and the upscattering cross section are plotted in Fig.~\ref{fig:UCNXection}. 

The anomalously high production of UCN in the $\beta$ phase can be explained by its large UCN production rate (as shown in Fig.~\ref{fig:UCNXection}), which primarily originates from exciting the soft diffuse mode clustered around a fixed $Q=1.3\mbox{\AA}^{-1}$ and extending from $\omega=0 \sim 8$ meV~\cite{Liu2011}. These modes are associated with the short range AF clusters present in the $\beta$-O$_2$ and should be related to the interesting cross-over behavior of the geometrically frustrated spins on a triangular lattice before the spins are fully ordered. The physics describing these excitations has not yet been fully formulated. Nevertheless, the inelastic scattering data provides the handle we need to proceed to understand the results of our UCN production experiment.

In a canonical superthermal source, where the loss mechanism is dominated by the upscattering loss, the ultimate achievable UCN density can be calculated from $\rho_{ucn}=P\times\tau=(N\sigma_{down}\phi_{CN})\times \frac{mfp}{v_{ucn}} = \frac{\sigma_{down}}{\sigma_{up}}\frac{\phi_{CN}}{v_{ucn}}$. It is the ratio of the downscattering to the upscattering cross sections that gives rise to the famous superthermal temperature dependence $\propto \exp(E/kT)$ (see the inset in Fig.~\ref{fig:UCNXection}). 
The superthermal behavior persists over a wide range of temperatures even when the mechanism of scattering changes abruptly between phases.
Note that this behavior is only evident with a large enough source, in which the UCN mfp is limited by the upscattering. 
With the reduction of upscattering by deep cooling, other sources of finite loss would become comparable in strength and finally limit the achievable gain. 

\begin{figure}[t]
\centering
\includegraphics[width=3.5 in]{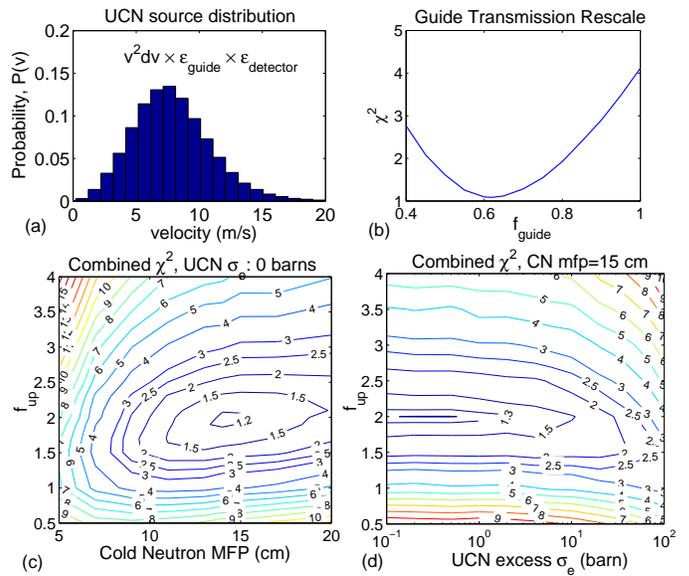}
\caption{\label{fig:Transmission} (a) The UCN velocity distribution inside the source used in the Monte-Carlo simulation. (b) Reduced $\chi^2$ and guide transmission rescale factor $f_{guide}$ to correct for the reduced transmission due to the installation of a UCN shutter, which affects the data sets collected using the small and large cells. Contour plots of the reduced $\chi^2$ with varying upscattering scale factor, $f_{up}$, and the CN mfp (c), and the UCN excess cross section (d).}
\end{figure}

Next, the UCN transport in the UCN guide system can be understood using Monte-Carlo simulations~\cite{Lavelle2010}. 
However, a UCN shutter was installed in the middle of the experiment leading to a reduction of the overall transport efficiency for the data set taken with small and large cells. Lacking sufficient data to isolate this effect, we introduce a free parameter, $f_{guide}$, to rescale the UCN signals for the small and large cells uniformly up, in order to compare with data taken using the medium sized cell. The UCN spectrum produced inside the source is described by the $v^2dv$ distribution, with the velocity $v$ extending beyond the critical velocity of UCN. The UCN guide serves as a spectrum filter, allowing the low velocity UCN (with energies below the Fermi potential of the guide, after gaining an energy boost of 87 neV exiting the oxygen source and entering the guide vacuum) to be transported with 100\% efficiency and filtering out the high velocity neutrons (very cold neutrons, VCN). However, because of the large angular acceptance and short length of our UCN guides, there remains a significant fraction of VCN reaching the detector. A reasonable neutron source distribution (derived from a Monte-Carlo simulation of UCN transport through the guide system) is presented in Fig.~\ref{fig:Transmission}a to account for these transport effects in our apparatus.   

Finally, the production of UCN inside the finite sized cell should not be spatially uniform because the CN flux is attenuated as it passes through the cell, mainly by elastic scattering. In our initial analysis, we scale the production rate, Fig.(~\ref{fig:UCNXection}), by an exponential decay function along the length of the cell, leaving the CN mfp as a free parameter to be determined by the best experimental fit. Note that this treatment assumes that the CN attenuates at the same rate independent of the energy. The model was later refined to include the evolution of the energy spectrum due to the strong Bragg scattering that is highly dependent on the CN wavelength.  

We build a model with a UCN velocity distribution according to Fig.~\ref{fig:Transmission}a, and the volume distribution according to the CN attenuation, using the temperature-dependent UCN production rate (Fig.~\ref{fig:UCNXection}). We then allow the individual source UCN to propagate uniformly through 4$\pi$ with the physics of UCN diffusion described by an excess elastic mfp to account for the material non-uniformity (Otherwise, the UCN elastic mfp is infinite in a uniform O$_2$ solid). The process of UCN diffusion also includes the energy-dependent upscattering loss and the nuclear absorption loss. The temperature dependence of density variation is included in the model. In addition to this excess elastic mfp, we add in additional 0.5~b and 2.5~b to the UCN elastic scattering in $\beta$-O$_2$ and $\gamma$-O$_2$ respectively, as suggested by the increased total cross-section measured with long-wavelength neutrons below the lowest Bragg cutoff. This additional scattering probably originates from paramagnetic scattering. The model assumes that the excess elastic scattering (due to the physical cracks) is fixed among all phases. It is probably well justified, considering that the change of the lattice structure and the molar volume from $\beta$ to $\alpha$ is minimal, and care were taken to avoid thermal shocks during the experiment. The CN attenuation is also treated the same in all phases, as the difference of the total cross section in $\alpha$ and $\beta$ is minuscule, due to their similar lattice structure and lattice constants (but it is not quite adequate to extend this assumption to the $\gamma$ phase).  

We then perform Monte-Carlo simulations using this basic model to calculate UCN yields from the 3 different cells, scanning through the full parameter space allowing variations of $f_{guide}$, CN mfp, and the upscattering normalization factor $f_{up}$ that translates the upscattering loss (in Fig.~\ref{fig:UCNXection}) into the absolute cross section. The reduced $\chi^2$ for each set of parameters was obtained by allowing the overall normalization, $N$, to be adjusted until the combined $\chi^2$ of the data from 3 cells,
\begin{equation}
\bar{\chi}^2 = \frac{1}{\sum_{cell,i}}\left[\sum_{cell,i}\frac{\left(S^{cell}_i(T_i)-N\times S^{cell}_{sim}(T_i)\right)^2}{\sigma_i^2+\sigma_{sim}^2}\right],
\end{equation}
is minimized for each temperature point independently (as $N$ contains no physical significance). The $cell$ index runs from small, medium, to large.  
Results of this study are shown in Fig.~\ref{fig:Transmission}. The best fit indicates that the transmission was reduced to about 62\% with the installation of an open UCN shutter, the upscattering scale factor is around 1.9, the CN mfp is about 14 cm, and the excess UCN elastic cross section is below 1 barn.  
The best fit with these optimum parameters to the experimental data is shown in Fig.~\ref{fig:BestFit}.
Note that not only the temperature dependence, but also the non-linear volume dependence of the UCN yield are reproduced in the simulation.


\begin{figure}[t]
\centering
\includegraphics[width=3.3 in]{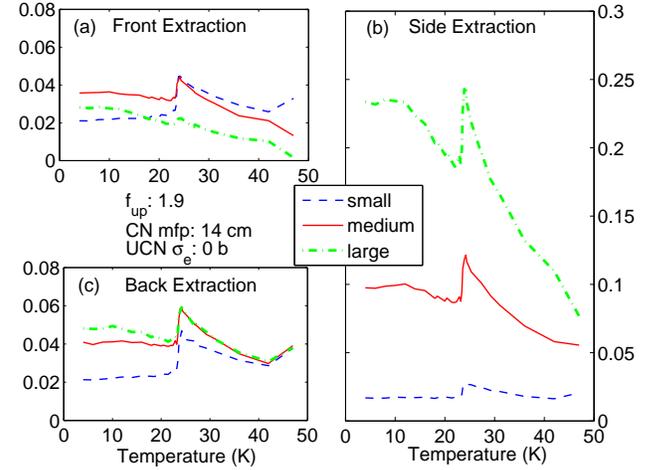}
\caption{\label{fig:Prediction} Simulated results of the UCN yield from s-O$_2$ extracted from (a) downstream of the CN flux (front extraction), (b) perpendicular to the CN flux (side extraction), and (c) upstream of the CN flux (back extraction).}
\end{figure}

The same simulation predicts different UCN yields for UCN extracted from different surfaces of our cylindrical cell (see Fig.~\ref{fig:Prediction}). In our experiment, the CN flux enters the cell horizontally from one surface and exits from the opposite surface. In what we shall term "front extraction", the UCN produced are extracted downstream of the CN flux. The UCN acceptance into the guide system is largest for UCN born in the volume slice close to the downstream window. On the other hand, the CN are significantly attenuated downstream, leading to a smaller amount of UCN produced in this volume. This is why the large cell does not produce the largest UCN yield in our front extraction configuration. The size of this effective volume, in which the UCN can be efficiently extracted, scales with the total mfp, combining effects of elastic scattering, upscattering and nuclear absorption. The expected superthermal temperature dependence is only evident when the UCN mfp is limited by the upscattering, and is smaller or comparable to the smallest dimension of the cell. In $\alpha$-O$_2$, both the production and upscattering are smaller, and thus the upscattering mfp is already much larger than the cell length, resulting in a very weak temperature dependence as shown in both the experimental data and the simulations. On the other hand, the number of UCN extracted from the upstream window (back extraction), and from the side (side extraction, including vertical extraction) increases monotonically with the increasing volume of s-O$_2$. Most noticeably, the expected temperature dependence from a superthermal source is restored in a large cell with UCN extracted from the side of the cell (into a UCN guide oriented perpendicular to the CN beam direction).   
This explains the seeming contradiction between the experimental data reported in ~\cite{Frei2010} and in this paper, as a purely geometrical effect.

\begin{figure}
\includegraphics[width=3.5 in]{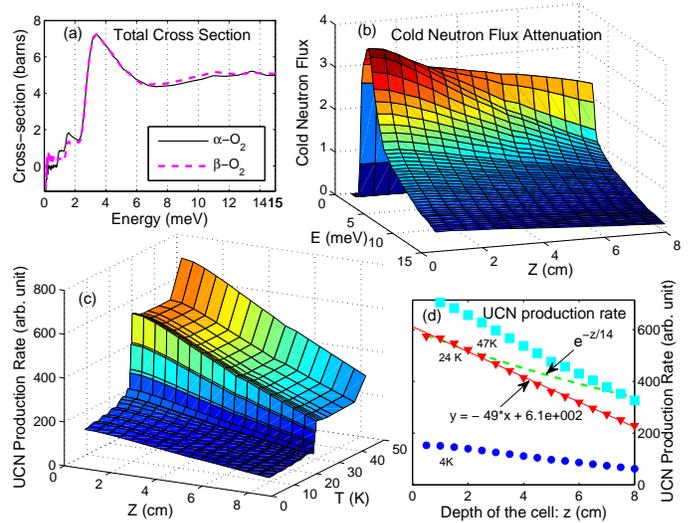}
\caption{\label{fig:CNTotal} (a) Total cross section of CN scattering in s-O$_2$. (b) Simulated results of the CN flux attenuation (along the depth of the cell $Z$) and the energy spectrum evolution. (c) UCN production rate weighted by the modified CN spectrum. (d) Linear fits to the reduced UCN production rate.}
\end{figure}

As the CN flux is attenuated in the solid, its energy spectrum also evolves as a result of the energy-dependent cross-section (see Fig.~\ref{fig:CNTotal}a). This modifies the spectrum-averaged UCN production rate used in the simulation discussed above. To study this effect, we developed another Monte-Carlo simulation (based on Geant4) for CN tracking, in which the CN elastically scatter isotropically with probabilities determined by the energy dependent total cross section, which is in fact dominantly elastic.   
Since the production of UCN in the source does not depend on the direction of the CN flux (incident or scattered), we record the usable flux at each point in space without taking the projection along any specific normal vector.  
The resulting flux attenuation over the depth of the cell, $Z$, and spectrum evolution are shown in Fig.~\ref{fig:CNTotal}b. With this information, we proceed to calculate the spectrum averaged UCN production rate at each volume slice along $Z$, for individual temperature points of interest (Fig.~\ref{fig:CNTotal}c). The reduction of the UCN production rate over the length of the cell can be fitted by a linear function (the first terms of the Taylor expansion of the exponential function). If we were to recast the reduction of the production rate as solely due to the CN flux attenuation, we arrive at a mfp of 12.5 cm, which is slightly smaller than the best value of 14 cm (obtained in the previous analysis where the CN mfp is left as a free parameter). This removes one free parameter required to perform the combined $\chi^2$ analysis described earlier. The revised analysis yields a larger $\chi^2$, 
leaving the best values of $f_{guide}$, $f_{up}$, and the excess UCN elastic cross section the same.
However, the constraint on the UCN elastic cross section is not tight enough to reach any definite conclusion about the limiting size of a practical UCN source made of s-O$_2$. 
The success of the fit validates the UCN production cross section extracted from the dynamic structure function.
We conclude that the UCN production in s-O$_2$ is highly enhanced by inelastic magnetic scattering, in particular, the large UCN production observed in the $\beta$ phase can be attributed to the excitations of a geometrically frustrated spin system.
This work was supported by NSF grant 0457219, 0758018.

\bibliography{O2_Vdep}

\end{document}